\documentclass[amssymb,prb,twocolumn,showpacs]{revtex4}
\usepackage{graphicx}
\usepackage{epsfig}
\newfont{\bl}{cmbxsl10 scaled\magstep1}
\begin{document}
\title{Voltage controlled spin injection in a (Ga,Mn)As/(Al,Ga)As Zener diode}
\author{P. Van Dorpe} \affiliation{IMEC, Kapeldreef 75, B-3001 Leuven, Belgium}
\email{pvandorp@imec.be}
\author{W. Van Roy} \affiliation{IMEC, Kapeldreef 75, B-3001 Leuven, Belgium}
\author{J. De Boeck} \affiliation{IMEC, Kapeldreef 75, B-3001 Leuven, Belgium}
\author{G. Borghs} \affiliation{IMEC, Kapeldreef 75, B-3001 Leuven, Belgium}
\author{P. Sankowski} \affiliation{Institute of Physics,
  Polish Academy of Sciences, al.~Lotnik\'ow 32/46, 02-668 Warszawa, Poland }
\author{P. Kacman} \affiliation{Institute of Physics,
  Polish Academy of Sciences, al.~Lotnik\'ow 32/46, 02-668 Warszawa, Poland }
\author{J.A. Majewski} \affiliation{Institute of Theoretical Physics,
Warsaw University, ul. Ho\.za 69, 00-681 Warszawa, Poland}
\author{T. Dietl} \affiliation{Institute of Physics,
  Polish Academy of Sciences and ERATO Semiconductor Spintronics
  Project, al.~Lotnik\'ow 32/46, 02-668 Warszawa, Poland \\
Institute of Theoretical Physics, Warsaw University, ul. Ho\.za 69, 00-681 Warszawa, Poland }

\date{\today}
\begin{abstract}
The spin polarization of the electron current in a $p$-(Ga,Mn)As-$n$-(Al,Ga)As-Zener tunnel diode, which is
embedded in a light-emitting diode, has been studied theoretically. A series of self-consistent simulations
determines the charge distribution, the band bending, and the current-voltage characteristics for the entire
structure. An empirical tight-binding model, together with the Landauer-B\"{u}ttiker theory of coherent
transport has been developed to study the current spin polarization. This dual approach allows to explain the
experimentally observed high magnitude and strong bias dependence of the current spin polarization.

\end{abstract}
\maketitle

Spin injection is one of the target applications of ferromagnetic semiconductors, which can serve as a natural
supply of highly spin-polarized carriers. In particular, the most intensively studied ferromagnetic
semiconductor (Ga,Mn)As can be grown epitaxially on GaAs and the carrier-mediated ferromagnetism in this
material provides an elegant way to control the ferromagnetic properties by tuning the hole
concentration.\cite{bib-1-ohno-01} In the design of spintronic devices, the p-type character of (Ga,Mn)As
introduces a disadvantage due the low hole spin lifetimes in GaAs. It has been shown that interband (Zener)
tunneling from valence band electrons of (Ga,Mn)As to the conduction band of GaAs is a way to circumvent this
disadvantage.\cite{bib-2-hoda-01,bib-3-johnston-halperin-02} Recently, an injected spin polarization up to 80\%
at 4.6~K has been demonstrated in a (Ga,Mn)As based spin-light emitting diode (LED) using Zener
tunneling.\cite{bib-4-van-dorpe} Moreover, the degree of injected spin polarization exhibits a strong dependence
on the applied bias. The spin polarization reaches its maximum just above the electroluminescence threshold and
decreases dramatically at higher bias. This effect is very interesting for spintronic applications, since it
provides a manner to control the degree of injected spin polarization with the applied voltage.

In this Rapid Communication, we analyze theoretically the transport in the spin-LED as a function of the applied
bias by means of self-consistent simulations. Furthermore, we compute the degree of current spin polarization at
the (Ga,Mn)As/GaAs-interface by combining an empirical tight-binding model with the Landauer-B\"uttiker theory
of coherent transport. We show that the decrease of the polarization with bias is caused by an increased
electron tunneling from the valence band of depleted (Ga,Mn)As and non-magnetic GaAs. This offers the
opportunity to tune the spin current polarization by an external electric field.

\begin{figure}
   \epsfig{file=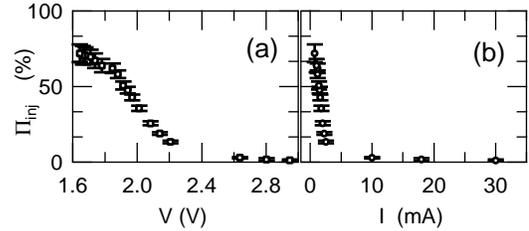,scale=1}
  \caption{ The degree of injected spin polarization measured at 4.6~K
    as a function of  the applied bias voltage (a) and the current (b) from Ref.~\onlinecite{bib-4-van-dorpe}.}
  \label{figure-1}
\end{figure}

The device considered here has the following structure: $p^+$ GaAs substrate / 200~nm $p$-Al$_{0.3}$Ga$_{0.7}$As
($2 \times 10^{18}$~cm$^{-3}$) / 100~nm p-GaAs ($2 \times 10^{18}$~cm$^{-3}$) / 60~nm n-Al$_{x}$Ga$_{1-x}$As ($1
\times 10^{17}$~cm$^{-3}$) / 30~nm n-Al$_{x}$Ga$_{1-x}$As ($1 \times 10^{18}$cm$^{-3}$) / 9~nm n-GaAs ($9\times
10^{18}$cm$^{-3}$) / 20~nm Ga$_{0.92}$Mn$_{0.08}$As, $i.e.$, the spin-LED from
Ref.~\onlinecite{bib-4-van-dorpe}. In this structure, the Al-concentration in the spin-drift region is
engineered together with the doping concentration in order to provide an effective barrier for the holes, such
that carrier generation due to impact ionization is eliminated at low bias. The measured\cite{bib-4-van-dorpe}
spin polarization is shown as a function of the applied bias voltage and current in Fig.~\ref{figure-1}(a) and
\ref{figure-1}(b), respectively. The spin polarization decreases dramatically with increasing bias.

\begin{figure}
  \epsfig{file=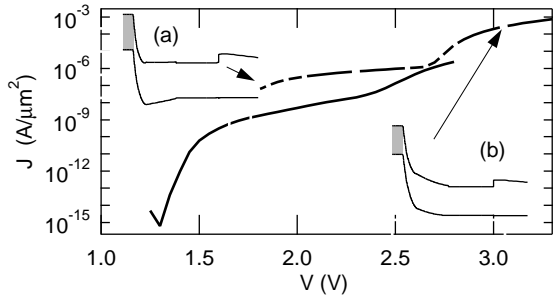,scale=1}
  \caption{The measured current (solid line) at 4.6~K as a function
    of the applied bias voltage (Ref.~\onlinecite{bib-4-van-dorpe})
    and the simulated current (dashed line) as a function of the applied bias voltage.
    The insets show the simulated conduction and valence band profiles at (a) 1.8 and (b) 3.0~V bias.
    The grey-colored area represents the (Ga,Mn)As injector.}
  \label{figure-2}
\end{figure}

We first model the charge transport in the device by means of a
series of self-consistent simulations of the entire LED under bias
using Medici,\cite{bib-11-medici} a semiconductor simulation tool
that allows self-consistent calculations of semiconductor
heterostructures taking into account band-to-band tunneling,
recombination and impact ionization. In these simulations the
(Ga,Mn)As region is treated as a heavily doped GaAs region. The
value of the doping level is chosen based on transport
measurements on samples grown using similar growth conditions on a
semi-insulating substrate. In the self-consistent simulations, the
interband tunneling is taken into account using the Kane model to
introduce a generation term.\cite{bib-6-kane-59} In the
simulations, the generation of carriers due to interband tunneling
is calculated assuming a completely filled valence band. Due to
the large Fermi energy ($E_F$) in (Ga,Mn)As, the carriers depart
from $E_F$ below the valence band maximum. This results in a small
error in the bias voltage where interband tunneling takes place,
but does not qualitatively alter the results of the calculations.
Impact ionization is treated in a post-processing mode, $i.e.$,
the carrier generation is calculated after the self-consistent
calculation of the band bending. For comparison, the simulated
current density at 40~K is shown together with the measured
current density at 5~K as a function of the applied bias voltage
on Fig.~\ref{figure-2}.
Due to well-known numerical convergence issues, we have been unable to simulate the structure at 5~K. There is a
good qualitative agreement between the simulated and the measured behavior, taking into account that the doping
levels and aluminum concentrations used in the simulations are nominal and can differ from reality. Between
1.6~V and 2.3-2.5~V the current is dominated by interband tunneling of electrons through the (Ga,Mn)As/GaAs
tunnel diode, while above 2.6~V the current is dominated by the holes, which can freely flow from the substrate
to the top contact.
\begin{figure}
\epsfig{file=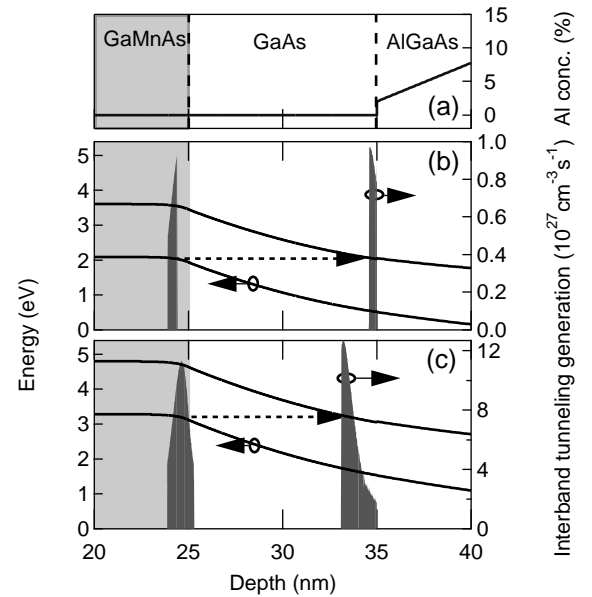,scale=1}
  \caption{The Al concentration (solid line) (a), the band diagram (solid lines)
    and the carrier generation (filled) due to interband
    tunneling (dotted line) as a function of the depth near the
    (Ga,Mn)As/GaAs tunnel diode at 1.8~V (b) and 3~V (c). The
    grey-coloured area represents the (Ga,Mn)As injector. The dotted
    arrows indicate the tunneling of valence electrons from (Ga,Mn)As
    to GaAs.}
  \label{figure-3}
\end{figure}
Figure \ref{figure-3} shows the band diagram (solid lines) of the
(Ga,Mn)As-(Al,Ga)As diode together with the Al concentration and
the carrier generation due to interband tunneling (filled black)
as a function of the depth near the (Ga,Mn)As/GaAs-interface at
1.8~V and at 3~V. There are two peaks present in the plotted
carrier generation, one in the (Ga,Mn)As-area, which shows the
generated holes in the (Ga,Mn)As valence band, and one in the GaAs
area that shows the generated electrons in the GaAs conduction
band. From this picture we can deduce that interband tunneling
takes place from the valence band of (Ga,Mn)As to the conduction
band of GaAs. When GaAs changes into (Al,Ga)As, the increasing Al
concentration causes a widening of the band gap and an abrupt
increase of the tunneling distance. This results in an
exponentially smaller tunneling probability and hence the number
of electrons that is generated in the (Al,Ga)As region is
negligible.  This means that only the part of the (Ga,Mn)As
valence band that overlaps with the GaAs conduction band can
participate in the tunneling process.  If we compare the low and
the high bias case we see that the area from where electrons
tunnel (the left peak in the carrier generation) is much bigger at
high bias than at low bias. When the voltage drop over the tunnel
diode increases, the part of the valence band of (Ga,Mn)As that
aligns with the conduction band of GaAs increases and hence the
tunneling originates from a wider region.
\begin{figure}
  \epsfig{file=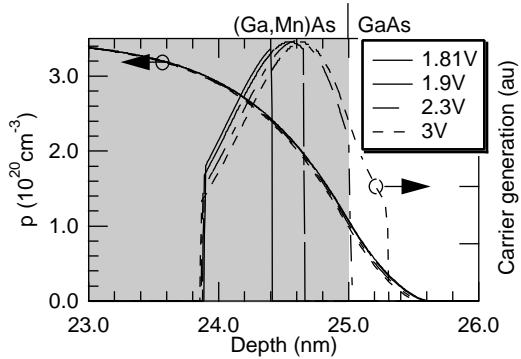,scale=1}
  \caption{The hole concentration (left) and the carrier generation
    due to interband tunneling (right) at different values of the
    applied bias voltage near the (Ga,Mn)As/GaAs interface. The
    grey-coloured area represents the (Ga,Mn)As injector.}
  \label{figure-4}
\end{figure}
In Fig. \ref{figure-4} we show the hole concentration and the
(normalized) carrier generation due to interband tunneling as a
function of the depth near the (Ga,Mn)As-GaAs interface for
different values of the applied bias voltage. The simulations show
that the hole concentration decreases in the bottom two nanometer
of the (Ga,Mn)As-layer due to the band bending in the
(Ga,Mn)As-GaAs p-n-diode. Increasing the voltage has a negligible
influence on the hole depletion near the interface.  However, we
do see a big change in the properties of the tunneling region. At
low bias, the tunneling electrons mainly depart from a region
where the hole concentration is smaller than, but close to the
bulk value, while at higher bias this region shifts and widens
such that electrons can tunnel from a region where the hole
concentration is much lower. At a somewhat higher bias ($>2.2$V)
we see that also electrons from the valence band of the
non-magnetic GaAs layer participate in the tunneling process. In
the experiment this bias will probably be lower due to the above
described small error in the interband tunneling model.

Apart from Zener tunneling, also impact ionization can contribute
to the generation of electrons in the GaAs conduction band. This
process generates unpolarized electron-hole pairs and hence
dilutes the injected spins and diminishes the measured spin
polarization. Impact ionization in this case is caused by holes
that flow from the substrate to the top contact and are heavily
accelerated by the strong electric field near the (Ga,Mn)As-GaAs
interface. However, the simulations indicate that the carrier
generation due to impact ionization only starts dominating at the
(second) hump in the IV-characteristics, where the valence band
reaches a flatband-situation (Fig.~\ref{figure-2}, inset (b)) and
the holes can flow freely from the substrate to the top contact.
Below this "hump" impact ionization can be neglected.

In parallel to self-consistent calculations that allow a detailed
understanding of the charge distribution, band bending, electron
and hole currents and their bias dependence, in order to study the
current spin polarization, we have developed a model of interband
tunneling, which combines an empirical tight-binding approach with
the Landauer-B\"{u}ttiker formalism.\cite{bib-8-fi-carlo-03} To
describe the band structure of GaAs we use the $sp^3d^5s^*$
tight-binding parametrization, with the spin-orbit coupling
included, proposed by Jancu {\em et al}.\cite{bib-9-jancu}  This
model reproduces correctly the effective masses and the band
structure of GaAs in the whole Brillouin zone, in agreement with
the results obtained by empirically corrected pseudopotential
method. The presence of Mn ions in GaMnAs is taken into account by
including the $sp$-$d$ exchange interactions within the virtual
crystal and mean-field approximations, with the values of the
exchange constants determined by the observed spin splittings of
the conduction and valence bands. Importantly, in contrast to the
standard $k\cdot p$ method, our model takes automatically into
account the Rashba and Dresselhaus terms, and is therefore
particularly well suited to describe interface phenomena such as
tunneling.

\begin{figure}
  \epsfig{file=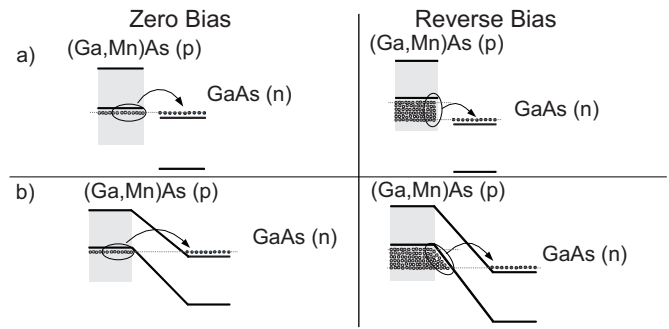,scale=1}
  \caption{Scheme of the structure used in the tight-binding calculations,
  at zero and at reverse bias without (a) and with the GaAs spacer (b).}
  \label{figure-6}
\end{figure}

Because of computational constraints in approaches involving transfer matrix formalism,\cite{bib-8-fi-carlo-03}
carrier transport along the whole device cannot be simulated. Therefore, we consider first the simplest
p-Ga$_{1-x}$Mn$_x$As/n-GaAs tunneling structure shown in Fig.~\ref{figure-6}(a).
Guided by previous theories of tunneling magnetoresistance (TMR), we
expect that while such model overestimates necessarily the tunneling
current, it can provide quantitative information on current spin
polarization that is determined by inter-band tunneling matrix
elements and degree of spin polarization in the ferromagnetic
electrode.

\begin{figure}
   \epsfig{file=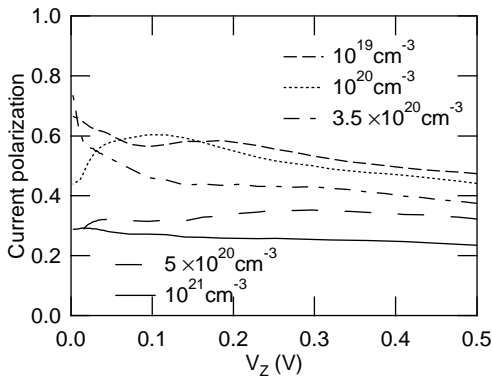,scale=1}
  \caption{The calculated bias dependence of the current polarization for
  different hole concentrations in the Ga$_{1-x}$Mn$_x$As-layer for structure (a) of Fig.~\ref{figure-6}.}
  \label{figure-7}
\end{figure}

In our computation, we assume that the magnetization vector is by
27$^{\circ}$ out of plane, as implied by experimental
conditions.\cite{bib-4-van-dorpe} We evaluate the spin current
polarization $P_j$ in respect to this direction. Furthermore, we
estimate that for a $T_{\mbox{\tiny{C}}}$ of 120~K and $x = 0.08$,
the expected hole concentration is of the order of $p = 3.5 \times
10^{20}$~cm$^{-3}$, as indicated by the experimental results in
Ref.~\onlinecite{bib-10-edmonds}. We also assume that the electron
concentration is $n = 10^{19}$~cm$^{-3}$. The dependence of $P_j$
on the hole concentration $p$ is depicted in Fig.~\ref{figure-7}.
Previous calculations have shown that the spin polarization of the
hole liquid decreases with $p$ in
Ga$_{1-x}$Mn$_x$As.\cite{bib-7-dietl-01} In this case, $P_j$, the
spin polarization of the tunneling electrons, shows a similar
behavior as function of $p$. This is naturally caused by the fact
that in the relevant range of $p$ and $x$, Ga$_{1-x}$Mn$_x$As is
not half-metallic, so that the Fermi energy is greater than the
spin-splitting. Surprisingly, however, a weak but opposite
behavior is seen at low $V_Z$. We assign this result to the fact
that the current polarization is determined not only by the
electron spin polarization at the Fermi energy but also by the
selection rules for transition probabilities. Actually, we know
that the mixing of the spin wave functions, which reduces the
tunneling matrix elements, is only important if the Fermi energy
is much smaller than the spin-orbit splitting of the valence band.
Accordingly, the reduction of $P_j$ by the spin-orbit coupling
decreases gradually with the hole density. Nevertheless, this
effect appears to be too weak to explain much smaller $P_j$ in the
non-annealed sample from Ref.~\onlinecite{bib-4-van-dorpe}.

\begin{figure}[t]
  \epsfig{file=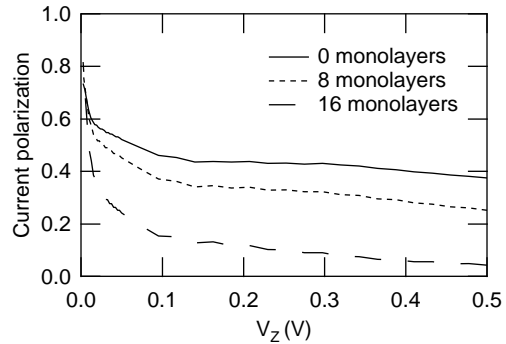,scale=1}
  \caption{The calculated bias dependence of the current polarization for different thicknesses of the spacer layer,
  for structure (b) of Fig.~\ref{figure-6}.}
  \label{figure-8}
\end{figure}

Next, to simulate more realistically the device, we consider the
structure consisting of Ga$_{1-x}$Mn$_x$As/GaAs/n-GaAs, where the
GaAs spacer has to be kept thinner than that of the n-GaAs
depleted layer implied by our self-consistent calculations. The
schematic of such structure is shown in Fig.~\ref{figure-6}(b).
Figure \ref{figure-8} shows $P_j$ as a function of the bias $V_Z$
of the Zener tunneling diode for various thicknesses $d$ of the
spacer layer.  We see that at low bias, $P_j$ depends weakly on
$d$ and is of the order of $0.7$, in perfect agreement with the
experimental results.  Interestingly, the drop of $P_j$ with
$V_Z$, is greater for larger $d$.  The strong dependence of $P_j$
on $V_Z$ is again in agreement with the experimental findings,
though a direct comparison is hampered by the fact that the exact
relation between $V_Z$ and the total bias $V$ applied to the
device is unknown. The dependencies of $P_j$ on $V_Z$ and $d$ can
be easily explained with the help of Fig.~\ref{figure-6}(b), which
shows that for $V_Z > 0$ the holes tunnel partly from the
non-magnetic GaAs.

In conclusion,  we have analyzed the spin polarization of the injected current in a (Ga,Mn)As-(Al,Ga)As spin-LED
by performing self-consistent simulations of the band bending and using tight-binding model together with the
Landauer-B\"{u}ttiker formalism for the calculation of the tunneling current. Our studies explain quantitatively
the large spin polarization of the injected current and its strong dependence on the applied voltage, as has
been recently observed in experiments. We ascribe the observed decrease of the spin polarization with the
applied bias to the enhanced tunneling from depleted (Ga,Mn)As and non-magnetic GaAs regions. This can provide a
path towards voltage controlled magnetic behavior. By choosing the right voltage drop over the tunnel diode, one
can switch between injection from a magnetized region to injection from a non-magnetized region.

P.V.D. acknowledges financial support from the I.W.T. (Flanders),
W.V.R. from the F.W.O. (Belgium), and P.S., P.K, and T.D. from the
Polish Ministry of Science, Grant PBZ-KBN-044/P03/2001. This work
is supported by the EC project FENIKS (G5RD-CT-2001-00535).


\begin{thebibliography}
  \frenchspacing
\bibitem{bib-1-ohno-01} H. Ohno, J. Cryst. Growth {\bf 251}, 285
  (2003).
\bibitem{bib-2-hoda-01} M. Kohda, Y. Ohno,  K. Takamura, F. Matsukura,
  and H. Ohno, Jpn. J. Appl. Phys.  {\bf40}, L1274 (2001).
\bibitem{bib-3-johnston-halperin-02} E. Johnston-Halperin, D. Lofgreen,
  R. K. Kawakami, D. K. Young, L. Coldren, A. C. Gossard, and
  D. D. Awschalom, Phys. Rev. B {\bf65}, 041306 (2002).
\bibitem{bib-4-van-dorpe} P. Van Dorpe, Z. Liu, W. Van Roy,
  V. F. Motsnyi, M. Sawicki, G. Borghs, and J. De Boeck,
  Appl. Phys. Lett. {\bf84}, 3495 (2004).
\bibitem{bib-11-medici} For details, see:
\url+http://www.synopsys.com+
\bibitem{bib-6-kane-59} E. O. Kane, J. Phys. Chem. Solids {\bf12},
  181 (1959).
\bibitem{bib-7-dietl-01} T. Dietl, H. Ohno, and F. Matsukura, Phys.
  Rev. B {\bf63}, 195205 (2001).
\bibitem{bib-8-fi-carlo-03} A. Di Carlo, Semicond. Sci. Technol.
  {\bf18}, R1 (2003), and the references therein.
\bibitem{bib-9-jancu} J.-M. Jancu, R. Scholz, F. Beltram, and
  F. Bassani, Phys. Rev. B {\bf57}, 6493 (1998).
\bibitem{bib-10-edmonds} K. W. Edmonds, K. Y. Wang, R. P. Campion,
  A. C. Neumann, C. T. Foxon, B. L. Gallagher, and P. C. Main,
  Appl. Phys. Lett. {\bf81}, 3010 (2002).

\end{thebibliography}
\end{document}